# Quantum tailoring of electronic properties in covalently functionalized graphene: application to ammonia gas detection

A. Dammak,[a] F. Raouafi,[a] A. Cavanna,[b] P. Rudolf,[c] D. di Caprio,[d] V. Sallet,[e] A. Madouri[b] and J. M. Jancu*[f]

Functionalized graphene offers great potential in the field of rapid detection of gases at room temperature. We performed first-principles calculations to study the suitability of 4-sulfobenzenediazonium salts (4SBD) as bandgap modifier in graphene. The signature of unpaired spins is evidenced near the Fermi level owing to the symmetry breaking of graphene sublattices. 4SBD-chemisorbed on graphene is found to be electronically sensitive to the presence of ammonia $NH_3$ with increasing gas concentration.

## 1 Introduction

Pristine graphene is a single two-dimensional layer of $sp^2$ bonded carbon atoms arranged in a honeycomb lattice.[1] In recent decades, it has been the subject of considerable study in most fields of science and engineering due to its unprecedented physical and chemical properties. In particular, graphene exhibits a large surface area, high electrical conductivity, optical transparency, exceptional thermal and mechanical behaviors, and ability to adsorb and desorb organic molecules.[2,3] It has been proposed for application in photovoltaics,[4] flexible electronics, gas sensors,[4,5] biosensors[6] and hydrogen storage.[7–10] Gas sensors are used in a wide range of applications for environmental monitoring, gas leak detection in industrial production, security, medical diagnostics, aircraft cabin air monitoring and food quality controls.[11,12] Solid state gas sensors represent a significant technological advancement in terms of low power consumption and specific active materials used for each type of application that determine their performance.[13,14] However, they suffer from long-term stability and some cannot operate at room temperature.[14] Chemically modified graphene is becoming one of the most versatile material for gas detection with superior performance in terms of electrical properties and sensitivity.[15–17] Both covalent and non-covalent approaches have been investigated for this purpose.[2,15–18] Covalent functionalization takes place when a functional group is covalently bonded to the surface. This enhances the aromatic character of graphene and improves long-term performance and structural stability compared to a non-covalent functionalization.[19,20] The generation of new covalent bonds in graphene can be easily obtained by free-radical addition using aryldiazonium salt chemistry.[19–23] For electron-withdrawing substituents on the aromatic ring, the overall process involves a transfer of a π-electron from graphene to the diazonium cation becoming an aryl radical after nitrogen removal. The resulting radical reacts directly with the graphene network forming an $sp^3$ covalent bond. In recent years, many research efforts have focused on controlling the density and spatial distribution of grafting sites.[17,24] Particularly noteworthy for covalent functionalization is the surface binding mechanism that can be customized depending on the substituent nature into the benzene ring.[25] Consequently, the lattice undergoes a non-uniform deformation, identical to a stress–strain response, which should be promising for engineering the electronic properties of functionalized graphene.[26,27] Covalently modified materials show a semiconductive behavior with finite bandgap and magnetic properties induced by the $sp^3$ hybridization.[17,28] On a theoretical level, previous *ab initio* calculations have been performed at different computational levels exploring a wide range of issues associated to the chemical configuration and binding energies of common organic molecules adsorbed on graphene.[29–36] In this context, the covalent functionalization of graphene with aromatic aziridines is found to open the bandgap from 0.3 eV to 0.5 eV.[35,36] However, very little is known about the functionalization of graphene with aryldiazonium salts.

In this work, we studied the electronic response of functionalized graphene with 4-sulfobenzenediazonium tetra-fluoroborate (4SBD). We performed first-principles pseudopotential plane wave calculations based on spin-

[a]*University of Carthage, IPEST, LPC2M, Route de Sidi Bou Saïd 2075 La Marsa, Tunisia*
[b]*C2N, University of Paris-Saclay, 10 Bd. Thomas Gobert, 91120 Palaiseau, France*
[c]*Surfaces and Thin Films Group, Zernike Institute for Advanced Materials, University of Groningen, The Netherlands*
[d]*IRCP, Chimie ParisTech, University of PSL, CNRS, 11 rue P. et M. Curie, 75005 Paris, France*
[e]*GEMaC, Université Versailles St-Quentin-en-Yvelines, France*
[f]*Univ Rennes, INSA Rennes, CNRS, Institut FOTON – UMR 6082, F-35000 Rennes, France. E-mail: jean-marc.jancu@insa-rennes.fr*

  

polarized density functional theory (DFT) and generalized gradient approximation (GGA) to model the electronic structure of 4SBD-modified graphene. The minimum energy path (MEP) governing the 4SBD dissociation within the kinetic process was determined by using the climbing image-nudged elastic band method (CI-NEB). The formation of local magnetic moments was addressed by the self-consistent calculation of the occupation for up and down spin states in the different orbitals of the complex. Graphene sublattices, hereafter referred to as *A* and *B*, are triangular lattices that span the honeycomb hexagon. We found that the sublattice imbalance induced by sulfobenzene adsorbates governs the bandgap properties and induces spin-polarized electronic states in the complex. The adsorption properties of ammonia molecules on modified graphene is investigated.

The article is organized as follows. In Section 2, we describe the computational approach. In Section 3, we present the adsorption properties of a single 4SBD molecule absorbed on graphene. In Section 4, we discuss the electronic and magnetic properties of functionalized graphene. In Section 5, we discuss the adsorption properties of $NH_3$ on 4SBD-decorated graphene. Summary in Section 5 concludes the article.

## 2 Computational details

Self-consistent DFT calculations were performed without spin–orbit coupling using the pseudopotential projector-augmented wave method implemented[37,38] in the Vienna *ab initio* simulation package (VASP 5.4.1) with an energy cutoff of 520 eV. Exchange–correlation (XC) effects were treated using the Perdew–Burke–Ernzerhof functional (PBE) generalized gradient approximation (GGA) functional.[38,39] A $(9 \times 9 \times 1)$ and $(12 \times 12 \times 1)$ $\Gamma$-centered *k*-points mesh was used to sample the Brillouin zone for geometry optimization and electronic structure respectively. We accounted for van der Waals (vdW) interactions in the dispersion-corrected DFT using the empirical correction approach.[13,14] Optimized lattice constants for pristine graphene are: $a = 2.46$ Å, $b = 2.46$ Å, $\alpha = \beta = 90°$ and $\gamma = 120°$, in excellent agreement with experimental results. Total energy calculations were performed using $4 \times 4$ and $6 \times 6$ supercells on the *xy*-plane. Since boundary conditions are applied in three dimensions, a 25 Å-thick vacuum region along the *z*-direction was included preventing any unphysical interaction between the periodic images of the slabs. All atomic sites were relaxed using the Hellman–Feynman theorem until forces are below of 0.01 eV Å$^{-2}$ with a total energy tolerance of $10^{-6}$ eV.

## 3 Adsorption of a single molecule on graphene

Physisorption (eqn (1)) and chemisorption (eqn (2)) correspond to the following reactions:

$$4SBD + Gr \rightarrow 4SBD–Gr \quad (1)$$

$$4SBD + Gr \rightarrow 4SB–Gr + N_2 \quad (2)$$

where Gr, represents pristine graphene, 4SB stands for the 4-sulfobenzene group and 4SBD for the 4-sulfobenzenediazonium salt. We calculated the adsorption energy from the energy difference between the interactive system (complex) and the non-interactive components. The 4SBD-physisorption energy is:

$$E_{ads}^{phys} = E(4SBD–Gr) - E(4SBD) - E(Gr) \quad (3)$$

$E$(4SBD–Gr) is the total energy of the complex, $E$(4SBD) the total energy of the free 4SBD molecule and $E$(Gr) the total energy of pristine graphene.

Similarly, the 4SBD-chemisorption energy is:

$$E_{ads}^{chem} = E(4SBD–Gr) - E(N_2) - E(Gr) - E(4SB) \quad (4)$$

$E$(4SB) is the total energy of the free 4SB molecule and $E(N_2)$ the total energy of nitrogen $N_2$.

Adsorption was studied using a $6 \times 6$ graphene supercell by positioning the molecule $N_2$ near the preferential sites shown in Fig. 1. We fixed a carbon atom as reference while the remaining atomic coordinates were relaxed to minimum energy.

### 3.1 Physisorption

Table 1 lists the nearest nitrogen-surface distances Sep for the physisorption sites. Geometry optimization was performed with an initial separation Sep = 2 Å, resulting in a strong overlap in electron density. The interplay between the short-range repulsive forces and van der Waals interaction leads to a relatively large separation Sep = 3.2 Å owing to the concave adsorption geometry displayed in Fig. 2a for the hollow site. We found a physisorption well depth of about 0.27 eV for all sites, which is similar to the binding energy of −0.38 eV for benzene adsorption on graphene.[40] The calculations are clearly site-independent thus demonstrating a non-localized process as expected for physical adsorption.[17]

### 3.2 Chemisorption

Chemisorption is site specific. Table 2 evidences that the 4SBD-adsorption at the T site is favored due to the equilibrium bond-distance $d_{ij} = 1.56$ Å corresponding to the interatomic bond-length of 1.54 Å in diamond. Indices *i* and *j* are referenced in Fig. 2b. The 1.8 eV chemisorption well depth is related to the electron-withdrawing character of the sulfo group ($SO_3H$) into the benzene ring[25] allowing an efficient binding of the molecule to the surface. In comparison, a DFT calculation gives $E_{ads}^{chem} =$

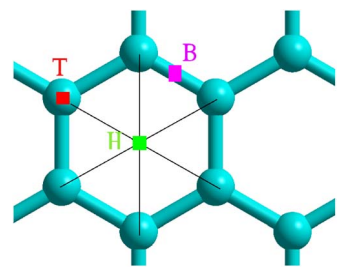

Fig. 1 Selected adsorption sites: bridge B, hollow H and top T.

Table 1　4SBD-physisorption parameters. Adsorption energy, $E_{ads}^{phys}$ associated to the separation distance Sep between graphene and nitrogen

|  | B | H | T |
| --- | --- | --- | --- |
| $E_{ads}^{phys}$ (eV) | −0.27 | −0.27 | −0.27 |
| Sep (Å) | 3.22 | 3.16 | 3.19 |

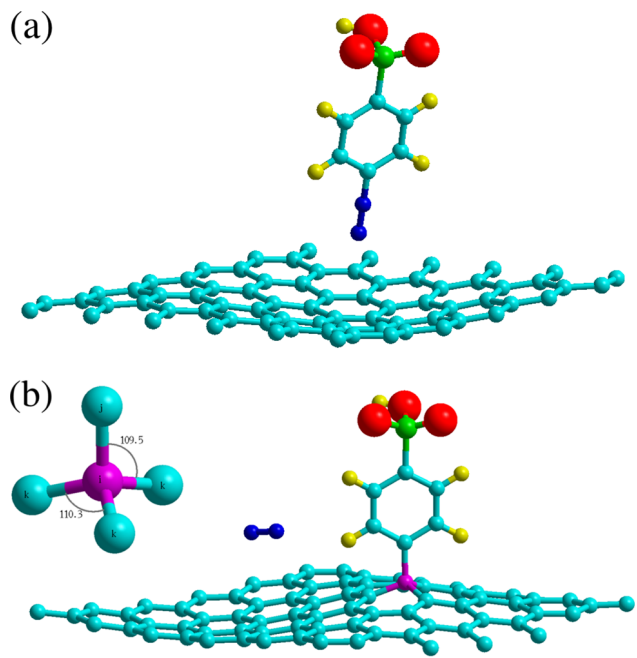

Fig. 2　Adsorption geometries of 4SBD diazonium salt on graphene. The top panel shows physisorption at the H site. The bottom panel shows chemisorption at the T site, the molecule is found non-tilted with respect to the graphene plane. Inset illustrates the sp$^3$-hybridized atom $i$ (in purple color) with its first-nearest neighbors $j$ and $k$. Atomic color code – carbon: turquoise; hydrogen: yellow; oxygen: red; nitrogen: blue; sulphur: green.

Table 2　Chemisorption parameters of a single 4SB molecule adsorbed on graphene. Adsorption energy $E_{ads}^{chem}$ and equilibrium binding distance $d_{ij}$ between atoms $i$ and $j$ as labeled in Fig. 2

|  | B | H | T |
| --- | --- | --- | --- |
| $E_{ads}^{chem}$ (eV) | −0.564 | −0.558 | −1.815 |
| $d_{ij}$ (Å) | 2.893 | 2.847 | 1.565 |

−0.99 eV for the phenyl radical $C_6H_5$ adsorbed on graphene.[40] Conversely, chemisorption at sites B and H is unlikely because the corresponding bond length $d_{ij}$ is twice of the covalent radius $r_{conv} = 0.76$ Å for sp$^3$ carbon according to the Pauling's equation. The 4SBD-optimized lattice is presented in Fig. 2b. Interestingly, atom $i$ brings out a tetrahedral geometry with the 109.5° near-ideal bond angles. Its nearest-neighbor atoms $k$ form the triangular basis and are seen to satisfy the threefold rotational symmetry along the $i$–$j$ axis. Atoms $k$ are therefore equivalent but their interatomic distance $d_{ik} = 1.51$ Å differs sensibly from the 1.42 Å bond-length in graphene. Chemisorption thus causes deformation where the surrounding of the grafting site experiences a tensile strain of 6%. Nevertheless, the bond-length variation turns out to be negligible in the nearest neighboring layer of atoms $k$, which stems from the ability of graphene to accommodate the adsorbate-induced strain beyond the grafting site.

It has been widely reported that diazonium salts spontaneously adsorb onto a wide range of materials including graphene.[22,41,42] A DFT simulation can shed light on this process by depositing the 4SB intermediate near the T site and the molecule $N_2$, for example, at 2 Å from the lattice. We observed that the radical steers adiabatically towards the adsorption site while $N_2$ bounces to 3.3 Å in height. In addition, the resulting adsorption parameters, $E_{ads}^{chem} = -1.34$ eV and $d_{ij} = 1.565$ Å, agree well with those of 4-SBD chemisorption.

### 3.3　Physisorption + dissociative chemisorption

The CI-NEB method allows to determine the minimum energy path (MEP) and transition state (TS) from an initial state (IS) to a final state (FS). Here, the MEP connects the 4SBD-physisorption (IS) to the chemisorbed product (FS). We estimated the MEP with a set of ten images shown in Fig. 3.

The dissociative energy barrier is calculated as the energy difference between the initial state IS (image 0) and the transition state TS (image 2) where the carbon–nitrogen bond is broken. We found an energy of 0.33 eV, which suggests an easy dissociation of nitrogen since it requires only a small amount of activation energy. Image 6 shows the covalent bonding of 4SB to the top site T. As seen in image 7, nitrogen diffuses easily on the surface since the adsorption energy decreases by 0.17 eV compared to image 6. The MEP gives a final state FS (image 9) less than the initial state IS (image 0). The chemical reaction therefore requires no activation energy defining a spontaneous grafting.

### 3.4　Charge density difference of 4SB on graphene

When the 4SB radical reacts with the graphene sheet, the complex experiences two modifications: a charge rearrangement and a distortion of its geometry (deformation effects). We calculated the DFT charge density difference as follows:

$$\Delta\rho(r) = \rho_{Gr-4SB}(r) - \rho_{Gr}(r) - \rho_{4SB}(r) \qquad (5)$$

where $\rho_{Gr-4SB}(r)$, $\rho_{Gr}(r)$ and $\rho_{4SB}(r)$ are the electronic density of the complex, pristine graphene and the free 4SB molecule respectively. Calculations were performed at fixed atomic positions after the structure optimization. Fig. 4 shows $\Delta\rho(r)$ where positive and negative charges are given with the same isosurface level. A significant variation in the charge density at the T site is observed due to the adsorption process. In particular, there is an enhanced interaction (accumulation of charge) between the radical and graphene by means of charge sharing between atoms $i$–$j$. Depletion occurs in the surrounding areas, which is consistent with the formation of a covalent bond. The charge accumulation near the oxygen atoms comes from the sulfo

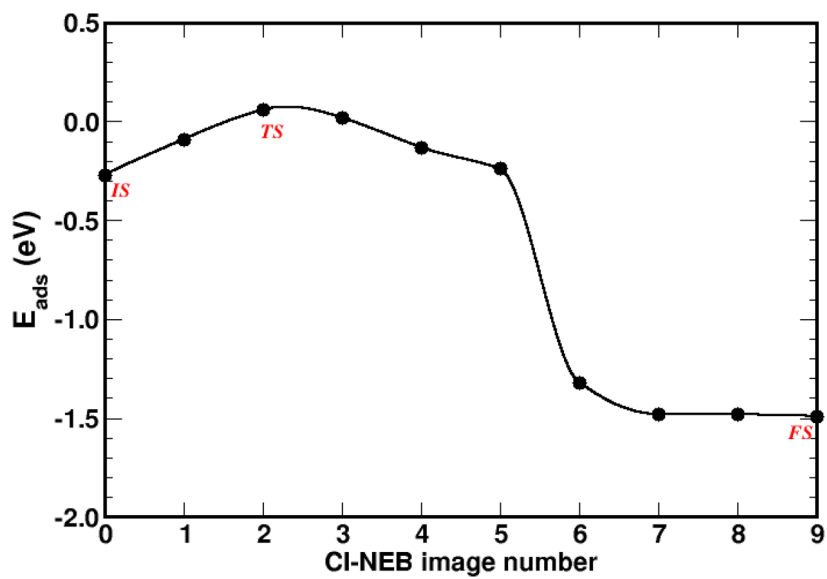

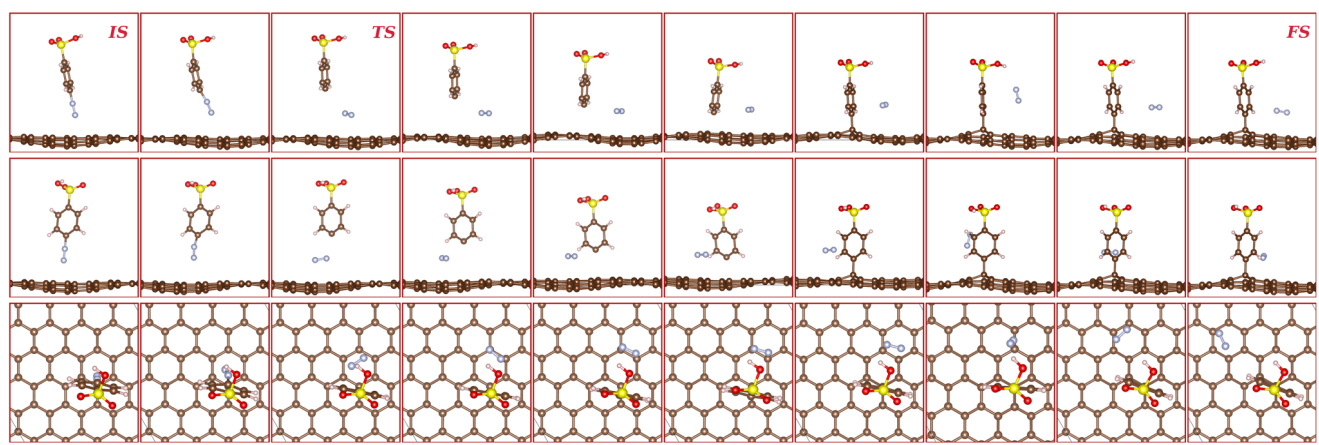

**Fig. 3** Climbing image-nudged elastic band (CI-NEB) pathway for known-on reaction in 4SBD-modified graphene. Top panel shows the minimum-energy pathway across a DFT potential energy surface as obtained from a CI-NEB calculation. The bottom panel shows ten representative CI-NEB images along the pathway from side, front, and top views. Image numeration increases from left to right. IS, FS and TS label the initial state, the final state and the transition state respectively.

group which strongly attracts the electron density from the benzene ring through inductive effects.

## 4 Spin-polarized density functional theory calculations

### 4.1 Electronic structure

We performed spin-polarized band structure calculations to account for the net magnetic moment of the complex. We used a 4 × 4 graphene supercell with a single 4SB molecule corresponding to a surface coverage rate of 25%. Fig. 5 shows the folded-band structure and total density of states (TDOS) of pristine graphene along the high symmetry directions. Band-folding can be observed at the $M$-point where the energy states are fourfold degenerate (with spin degeneracy) compared to the second-order degeneracy in the primitive cell (2 atoms).

Many papers have been published on graphene properties[43] focusing on the bonding (lower-energy valence-band) and antibonding (higher-energy conduction-band) states, referred $\pi$ and $\pi^*$ respectively. As illustrated in Fig. 5, the bandgap between the $\pi$ and $\pi^*$ bands closes at the Fermi level $E_F$ lying at the Dirac energy $E_D$ on the $K$-points. Near the Dirac point $E_D$, the energy-band dispersion increases linearly with the wave vector $k$ and the $\pi$-electrons are generally imaged as "massless Dirac fermion". Therefore, neglecting thermal excitations, the intrinsic charge carrier density is strictly zero in the pristine structure. Identifying a suitable approach to control the Fermi level and thus dissociate $E_F$ from $E_D$ is of crucial importance to overcome the low carrier density in graphene. Highly p-doped graphene can be obtained by introducing strong electron-withdrawing substituents into the aromatic ring.[44] Such a trend is evidenced for 4SBD-chemisorbed on graphene with a Fermi level lowered by 40 meV compared to the Dirac point $E_D$.

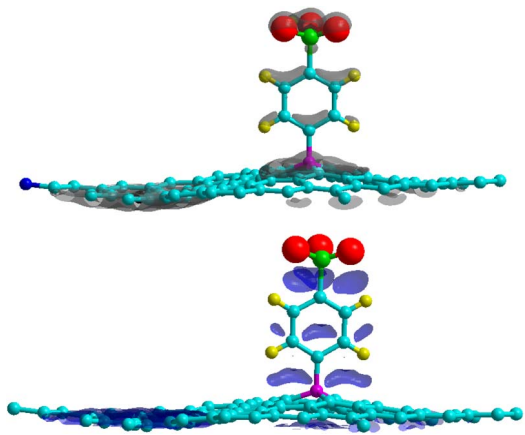

**Fig. 4** Charge density differences for 4SBD-adsorption on a 6 × 6 graphene supercell for electron gain (top panel) and depletion (bottom panel). Grey (blue) indicates regions of charge gain (loss). An isovalue of ±0.003 e Å$^{-3}$ is considered.

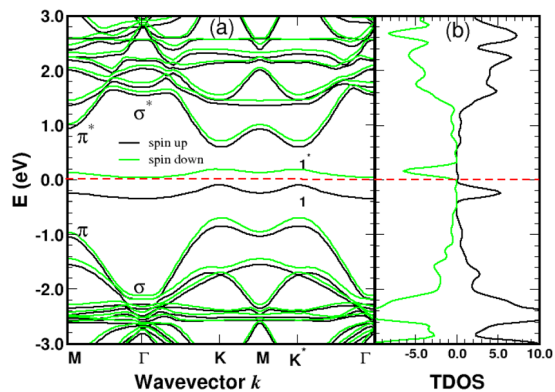

**Fig. 6** Spin polarized band-structure (a) and total density of states in arbitrary units (b) of a single 4SB molecule adsorbed on graphene. The σ and π-like states deriving from the pristine structure are indicated near the $\Gamma$ and $K$-points respectively. The spin impurity states related to the sp$^3$ hybridization are labeled 1 and 1*. The Fermi level of the complex is the zero-energy reference.

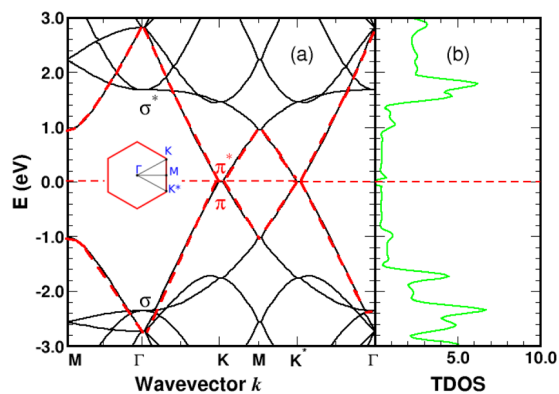

**Fig. 5** Graphene band-structure near the Dirac energy (a) and corresponding total density of states in arbitrary units (b) along the symmetry lines $\Gamma$–$K$–$M$. The π bands are marked in red. The Fermi level is the zero-energy reference. Inset shows the irreducible Brillouin zone with the two inequivalent $K$-points around which the energy dispersion becomes linear.

The spin-polarized bands of the complex and partial density of states (PDOS) of selected atoms are illustrated in Fig. 6 and 7. The calculations were performed without spin–orbit coupling (real and spin spaces are decoupled) leading to two independent sets of majority and minority spin bands. Clearly, the π-like bands, dominated by the p$_z$ orbitals of graphene, are very sensitive to the local chemical environment induced by the 4SB adsorbate. At the $K$-points, band-degeneracy of the π states in the primitive cell is lifted, determining the valence-band-maximum (VBM) and conduction-band-minimum (CBM). Therefore, the bandgap has a direct character allowing an optical excitation of 1.45 eV. Note that the 6-fold symmetry is broken in the complex because of non-uniform strain and the $K$-points become inequivalent. Interestingly, the energy-band dispersion at the band edges is parabolic and consequently the effective-mass approximation is valid. It should be noted that the σ-like states compare well with their equivalent (sp$^2$ hybridized) in Fig. 5 and the small energy differences are related to the lattice relaxation-induced lifting of the σ-degeneracy evidenced in Fig. 6. Therefore, the s, p$_x$ and p$_y$ orbitals of each carbon in graphene are weakly perturbed by the adsorbate-induced potential. Finally, the Bloch states derived from the highest occupied molecular orbital (HOMO) and the lowest unoccupied molecular orbital (LUMO) of 4SB are found to be weakly dispersive in the Brillouin zone and lying in energy at −2.04 eV and 1.63 eV respectively.

### 4.2 π-Magnetism

Graphene is non-magnetic with negligible spin–orbit coupling, which should make it suitable for spin-polarized applications.

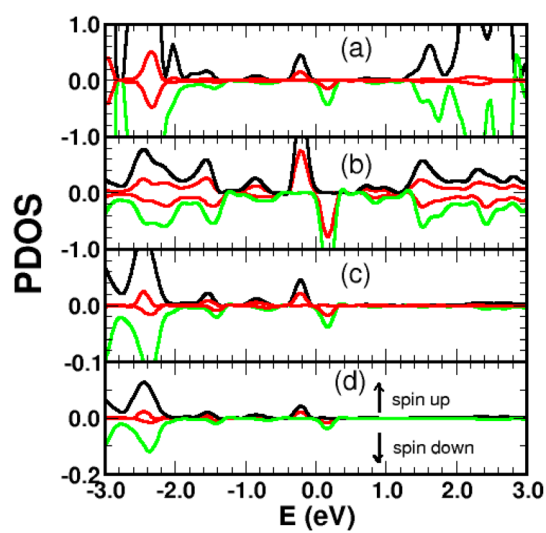

**Fig. 7** Spin-polarized projected density of states (PDOS) in arbitrary units for a single 4SB molecule chemisorbed on graphene. (a) 4SB molecule; (b) first-shell neighbors $k$; (c) atom $j$; (d) atom $i$. The indices $i$, $j$, $k$ are referenced in Fig. 2. The p$_z$-orbital contributions are shown in red.

Fig. 6a and b show that the chemical functionalization leads to two midgap impurity states near the Fermi energy defined by an occupied spin-up (marked 1) and an unoccupied spin-down (marked 1*) character. The unpaired electron 1 derives from the p$_z$ orbitals with a PDOS mainly localized around the grafting site as seen in Fig. 7b and c. Of note for graphene spintronics is the large spin separation of 420 meV in the TDOS peaks associated to the states 1 and 1* (Fig. 6b). Accordingly, we calculated a total spin magnetic moment $\mu_S = 0.98\mu_B$ that results from the sublattice imbalance in line with the following considerations. $\mu_S$ can be decomposed into the site-projected local spin moments as: $-0.01\mu_B$ for the reference atom $i$ (A-site), $0.51\mu_B$ for the first-shell neighbors $k$ (three B-sites), $-0.28\mu_B$ for the second-shell neighbors (six A-sites), $0.42\mu_B$ for the third-shell neighbors (six B-sites) and $0.31\mu_B$ for the fourth-shell neighbors (six B-sites). The more distant layers and the 4SB molecule contribute to a lesser extent. These values suggest that the long-range magnetic interaction mediated by direct exchange is favored. Therefore, the functionalization of graphene with 4SBD gives an antiferromagnetic character due to opposite signs on the A and B sublattices where their uncompensated local spin moments lead to the finite value of $\mu_S$. Our findings are in agreement with Elton's DFT calculations[34] performed on a wide range of adsorbates on graphene. It is worth noting that the π-magnetism induced by one single-carbon vacancy features a similar electronic band-diagram with magnetic states lying at midgap.[45] Further insight can be provided from the band-structure decomposition into a bonding contribution due to adsorbate–surface interactions and a mechanical contribution due to structural relaxation. The second one, displayed in Fig. 8, represents the electronic structure of strained graphene upon chemisorption. The energy-band dispersion is broadly similar to that calculated for pristine graphene.

There are no magnetic quantum states and lattice relaxation lifts all degeneracies from the primitive cell, resulting in a 150 meV bandgap at the $K$-points. In Conclusion, chemical bonding governs the electronic and magnetic properties of 4SBD-chemisorbed on graphene related to bandgap modification.

### 4.3 Coverage-dependent adsorption

We studied the properties for a coverage of $N_{SB}$ molecules of sulfobenzene chemisorbed on a 4 × 4 graphene supercell. A functionalization ratio of 50% ($N_{SB} = 2$) to 100% ($N_{SB} = 4$) is considered by placing one 4SB molecule above a carbon atom on the $B$-sublattice and the others 4SB above the carbon atoms on the $A$-sublattice as illustrated in Fig. 9. Additional studies of configuration effects are certainly appropriate but require detailed chemisorption modeling, which is beyond the scope of this article. The DFT adsorption energy per molecule shown in Fig. 10a is defined by:

$$E_{ads} = E(4SB–Gr) - E(Gr) - N_{SB}E(4SB) \quad (6)$$

We observed in the calculations that the bond-length $d_{ij}$ associated with a given $E_{ads}$ are all equivalent between the grafting sites at approximately 0.005 Å. Fig. 9b addresses $N_{SB} = 2$ where the two molecules are separated from each other by the fifth-nearest neighboring shell. Therefore, no overlapping orbitals or adsorbate–adsorbate interactions should alter the binding properties of the complex. Moreover, we calculated an interatomic distance $d_{ij} = 1.58$ Å reflecting an efficient chemisorption process with the bonding of the two molecules at the surface. As a result, the Fermi level $E_F$ and the binding energy shown in Fig. 10b and c increases in absolute value from $N_{SB} = 1$ to $N_{SB} = 2$ due to the enforced surface–adsorbate interactions. Very interestingly, the band-structure evidences in Fig. 11 the long-range ordered properties related to the equipartition of the SB molecules on the $A$ and $B$ sublattices for the layout $N_{SB} = 2$. Particular emphasis is placed on the suppression of the magnetic spin states leading to $\mu_S = 0$. This result is in agreement with Lieb's theorem,[46] where any imbalance in the number of absorption sites between the two sublattices leads to a magnetic ground state. Hydrogen chemisorption experiments on graphene also confirm finite or zero value of the total spin magnetic moment depending on the relative position of adsorbates on the two sublattices.[47] Moreover, the bandgap strongly decreases from $N_{SB} = 1$ to $N_{SB} = 2$ (Fig. 10c), which is a typical behavior of ordering as already observed in ordered semiconductor alloys.[48] Finally, we will note on the Fig. 11 the position of the Fermi level near the valence-band maximum occurring between the $\Gamma$ and $K^*$-points. Therefore, the electrical conductivity of the complex is expected to increase compared to $N_{SB} = 1$ (Fig. 6).

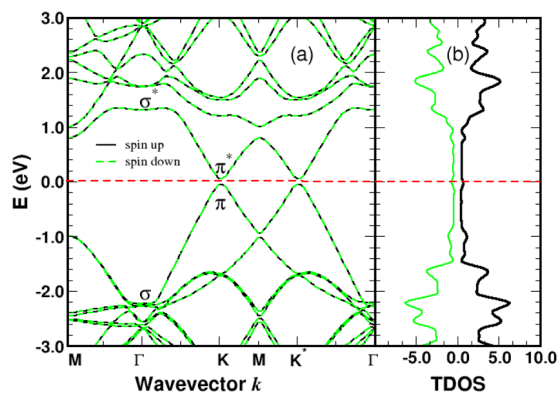

Fig. 8 Electronic band-structure (a) and density of states in arbitrary units (b) of strained graphene with the relaxed atomic positions due to the chemisorption process. The Fermi level of strained graphene is the zero-energy reference.

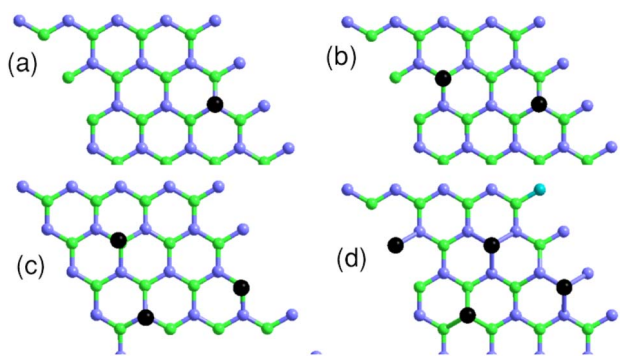

Fig. 9 Top view of the 4 × 4 graphene supercell with the sulfobenzene molecules indicated by black dots. (a) $N_{SB} = 1$; (b) $N_{SB} = 2$; (c) $N_{SB} = 3$; (d) $N_{SB} = 4$. Light-blue and green colors label the $A$ and $B$ sublattices respectively.

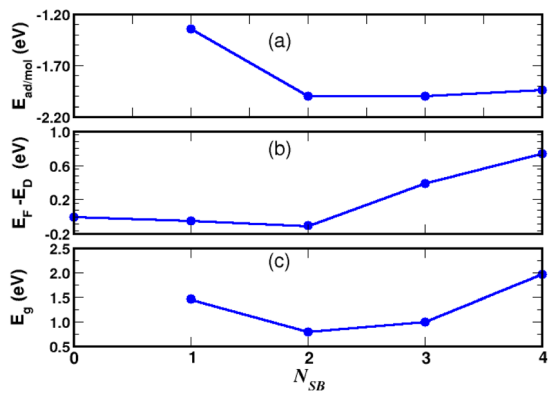

Fig. 10 Chemisorption of $N_{SB}$ molecules on graphene. Site occupancy is given in Fig. 9. Adsorption energy (a), Fermi level with respect to the Dirac energy $E_D$ of pristine graphene (b), and VBM-to-CBM bandgap (c).

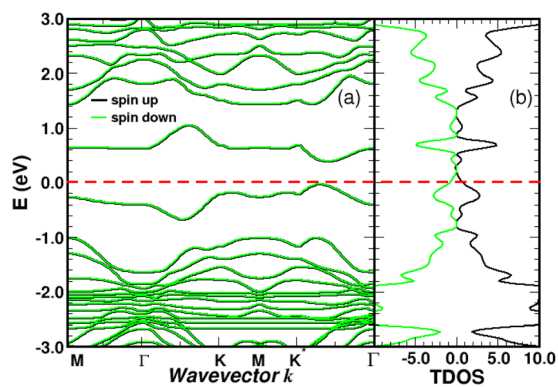

Fig. 11 Band structure (a) and density of states in arbitrary units (b) for two molecules of sulfobenzene adsorbed on graphene. 4SB-site occupancies are given in Fig. 9b. The Fermi level of the complex is the zero-energy reference.

For $N_{SB} = 3$ and $N_{SB} = 4$, the lateral or repulsive interactions between the adsorbates[49] are evidenced by the simulation of a stable binding energy with increasing surface coverage. A bond-length of $d_{ij} = 1.59$ Å is found in the two coverages and still compares well with the bulk diamond interatomic distance of 1.54 Å. Therefore, a high functionalization coverage with $N_{SB} = 4$ is a plausible scenario for 4SBD-chemisorbed on graphene and we used as a structure model in the following section. We would like also to point out that an ordered structure at high surface coverage should decrease the lateral interactions since ordering phenomena during atomic relaxation lead to a better accommodation of neighboring adsorbates. Finally, the spin-polarized electronic band-structure for $N_{SB} = 4$ is illustrated in Fig. 12. The modified material exhibits a strong semiconductor character with a VBM-to-CBM bandgap of 1.96 eV occurring at the $\Gamma$-point and the associated energy-band dispersion is parabolic. Moreover, a net spin magnetic moment of $\mu_S = 2\mu_B$ is calculated, deriving from the unpaired electron states labeled 1 and 2 in Fig. 12. This result is in agreement with Lieb's theorem and the imbalance of two 4SB adsorbates between the $A$ and $B$ sublattices. These two singly states lying below the Fermi energy should allow the transfer of two electrons from graphene to them, thus increasing the p-doping of the complex compared to $N_{SB} = 1$ (Fig. 6).

## 5 Ammonia adsorption on 4SBD-decorated graphene

Theoretical and experimental findings showed that the binding energy of $NH_3$ on pristine graphene lies in the physisorption values around 30–60 meV, with the interactions being dominated by the van der Waals forces.[50–52] Here, we demonstrated that the process can be enhanced with the use of 4SBD-decorated graphene ($N_{SB} = 4$) due to the strong orbital mixing between ammonia and the sulfo group, thus demonstrating a chemical contribution to the adsorption bond. Fig. 13 illustrates the charge density changes after adsorption of a single ammonia on the complex. An isovalue of $\pm 0.003$ e Å$^{-3}$ was used to display the accumulation of charge density within the interaction/bonding regions. Adsorption is evidenced by the accumulation of electron density in the vicinity of the oxygen regions and depletion near hydrogen atoms. The adsorption properties was investigated as function of the number $N_{NH_3}$ of ammonia molecules upon the complex. The optimized geometry is given in Fig. 14 for $N_{NH_3} = 4$ and displays some differences compared to the atomic structure of 4SBD-modified graphene without the gas. There is a formation of a single ammonium $NH_4^+$ due to the elimination of the proton from the sulfo group. Calculations showed that the $SO_3^-$ fragment shortens the S–O single bond-length by 0.02 Å compared to that of sulfonic acid while the S=O double bond-lengths increase slightly by 0.005 Å. Moreover, the hydrogen atoms belonging to the other three $NH_3$ molecules are located approximately at 2.4 Å from the substituent.

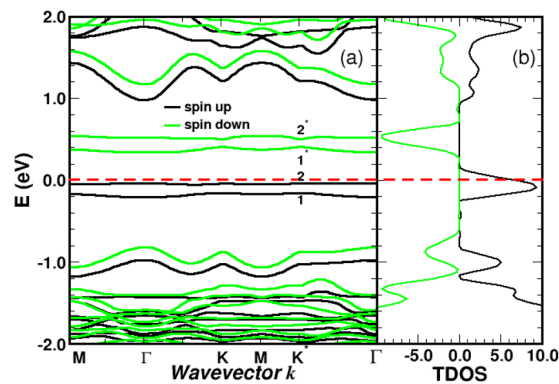

Fig. 12 Band structure (a) and density of states in arbitrary units (b) for four molecules of sulfobenzene adsorbed on graphene. 4SB-site occupancies are given in Fig. 9d. The spin impurity states deriving from the sp$^3$ hybridization are labeled 1, 2, 1*, and 2*. The Fermi level is the zero-energy reference.

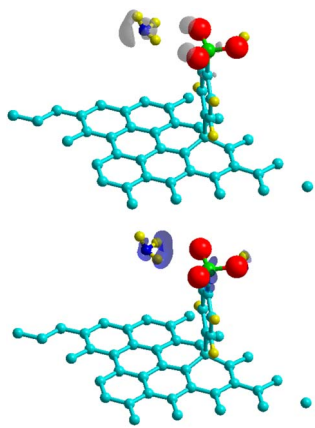

Fig. 13  Charge density differences for $NH_3$-adsorption on a 4 × 4 4SBD-decorated graphene for electron gain (top panel) and depletion (bottom panel). Grey (blue) indicates regions of charge gain (loss).

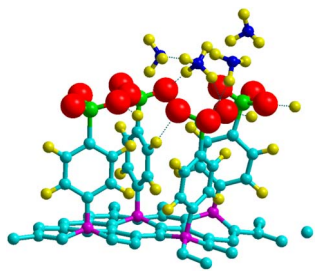

Fig. 14  Adsorption geometry for four $NH_3$ molecules adsorbed on 4SBD-decorated graphene.

Turning now to the electronic structure for $N_{NH_3} = 4$ (Fig. 15 and 16) we note that the band energies and TDOS are similar to those displayed in Fig. 12. Therefore, the gas interacts only weakly with the complex, which is suited for applications in sensors. Differences are related to the emergence of well-definite ammonia-derived TDOS peaks (located at −0.44 eV, −1.15 eV and −1.75 eV) with a sharp linewidth (0.14 eV full width at half maximum) that give considerable strength for physisorption. These adsorbate-induced peaks results from hybridization between the molecular orbitals of ammonia and the sulfonic acid and from the tendency of the gas to form hydrogen bonds with oxygen atoms, i.e. the sulfo group is a hydrogen bond acceptor. This analysis is supported by the calculation of the adsorption energies shown in Fig. 17a and defined as:

$$E_{ads/mol} = \frac{1}{N_{NH_3}} \left( E(Gr_{modified} - gas) - E(Gr_{modified}) - E(gas) \right)$$

(7)

where $E(Gr_{modified}$–gas) is the total energy of the complex with the adsorbed molecules, $E(Gr_{modified})$ is the total energy of 4SBD-decorated graphene and $E(gas)$ is the total energy of the ammonia gas. The physisorption range is found between 100 meV and 200 meV associated with a stable bandgap for the different gas concentrations (Fig. 17b). We would like to point out that the calculated value of the interaction energy between $NH_3$ and free 4-sulfo-benzene amounts to −0.07 eV. This value is of the same order of magnitude as van der Waals interactions (10–100 meV), whereas the physisorption energy of $NH_3$ on functionalized graphene is around −0.21 eV in our calculations. This last result is about the same order of magnitude as hydrogen bonds (200–400 meV). $NH_3$ is therefore more strongly adsorbed on the functionalized graphene than on the 4SBD molecule alone. It should be also noted that the binding energy decreases rapidly with increasing number of ammonia molecules (Fig. 17a), which is interesting to promote high-efficiency desorption.

Moreover, the Fermi level (Fig. 17c) increases significantly in the complex as function of the number $N_{NH_3}$ and such a process can be explained by the action of ammonia as charge donors.[51] Finally, we would like to point out that the electronic structure upon adsorption processes as a whole, involving the 4SBD diazonium salt and ammonia molecules, induces an opposite type of doping in the graphene surface (p-type with 4SBD and n-type with NH$_3$). Consequently, the electrical conductivity is expected to decrease with increasing gas concentration.

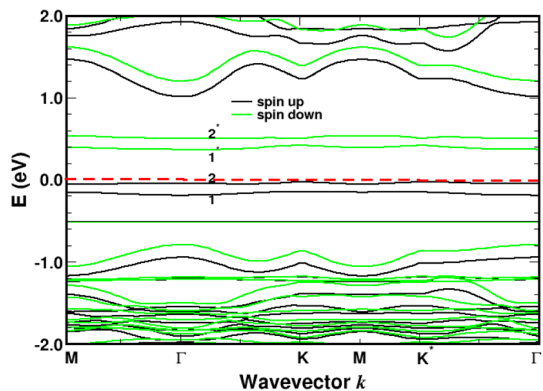

Fig. 15  Band structure for four molecules of $NH_3$ adsorbed on 4SBD-modified graphene with a surface-coverage of 100%. The $sp^3$-like defects are labeled 1, 2, 1*, 2*. The Fermi level is the zero-energy reference.

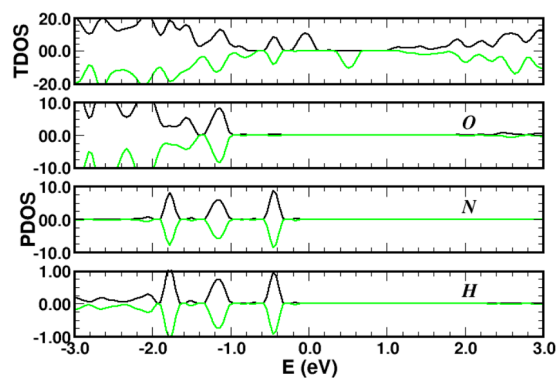

Fig. 16  Total (top panels) and atomic-projected density of states (lower panels) on oxygen (O), hydrogen (H) and nitrogen (N) for four $NH_3$ molecules adsorbed on 4SBD-decorated graphene, in arbitrary units.

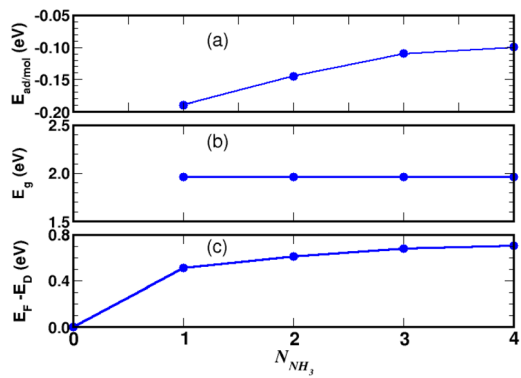

Fig. 17 NH$_3$-adsorption on 4SBD-decorated graphene as function of the number $N_{NH_3}$ of ammonia molecules. Binding energy (a), VBM-to-CBM bandgap (b), Fermi level with respect to the Dirac energy $E_D$ of pristine graphene (c).

## 6 Conclusion

In summary, we investigated from DFT the chemisorption properties and electronic structure of graphene functionalized with sulfobenzenediazonium salts. The functional group forms a stable molecular system and interacts strongly with graphene even at high coverage giving an efficient chemisorption process. The electronic properties of 4SBD-modified graphene are governed by the lattice imbalance due to the difference in the number of adsorbed atoms in the individual sublattices. Calculations also revealed spin signature of individual sp$^3$-type defects near the Fermi level making the magnetic properties directly accessible in experiment. The complex is found sensitive to the adsorption of ammonia and demonstrate efficient desorption activity. Further modeling concerns the dependence of the electronic structure on the different adsorbate configurations related to the surface properties of graphene.

## Author contributions

Conceptualization, V. S., A. C. and A. M.; methodology, F. R. and J. M. J.; software, D. d. C., A. D. and F. R.; formal analysis, F. R. and J. M. J.; investigation, A. D., D. d. C. and F. R.; resources, D. d. C.; writing – original draft preparation, A. D. and F. R.; writing – review and editing, P. R., F. R. and J. M. J.; supervision, F. R. and J. M. J.; All authors have read and agreed to the published version of the manuscript.

## Conflicts of interest

There are no conflicts to declare.

## Acknowledgements

We thank the financial support of the partenariat Hubert Curien Utique PHC UTIQUE 17G1207 from the French ministry Ministère de l'Europe et des Affaires Etrangères and the Tunisian ministry Ministère de l'Enseignement Supérieur et de la Recherche Scientifique.

## Notes and references